# Thermal with Electronic Excitation for the Unidirectional Rotation of a Molecule on Surface

*Kwan Ho Au-Yeung[1], Suchetana Sarkar[1], Tim Kühne[1], Oumaima Aiboudi[2,3], Dmitry A. Ryndyk[4,5], Roberto Robles[6], Franziska Lissel[2,3], Nicolas Lorente[6,7], Christian Joachim[8], Francesca Moresco[1*]*

[1]Center for Advancing Electronics Dresden, TU Dresden, 01062 Dresden, Germany

[2]Leibniz-Institut für Polymerforschung Dresden e.V., 01069 Dresden, Germany

[3]Faculty of Chemistry and Food Chemistry, TU Dresden, 01062 Dresden, Germany

[4]Institute for Materials Science, TU Dresden, 01062 Dresden, Germany

[5]Theoretical Chemistry, TU Dresden, 01062 Dresden, Germany

[6]Centro de Física de Materiales CFM/MPC (CSIC-UPV/EHU), 20018 Donostia-San Sebastián, Spain

[7]Donostia international physics center, 20018 Donostia-San Sebastián, Spain

[8]GNS & MANA Satellite, CEMES, CNRS, 29 rue J. Marvig, 31055 Toulouse, France

ABSTRACT. Exploring the limits of the microscopic reversibility principle, we investigated the interplay between thermal and electron tunneling excitations for the unidirectional rotation of a molecule-rotor on the Au(111) surface. We identified a range of moderate voltages and temperatures where heating the surface enhances the unidirectional rotational rate of a chemisorbed DMNI-P rotor. At higher voltage, inelastic tunneling effects dominate while at higher temperature the process becomes stochastic. At each electron transfer event during tunneling, the quantum mixing of ground and excited electronic states brings part of the surface thermal energy in the excited electronic states of the molecule-rotor. Thermal energy contributes therefore to the semi-classical unidirectional rotation without contradicting the microscopic reversibility principle.

INTRODUCTION

The unidirectional rotation of a single-molecule rotor adsorbed on a surface cannot be induced thermally without breaking the microscopic reversibility principle.[1] According to the energy equipartition theorem, the thermal energy coming from the surface is equally distributed to all accessible mechanical degrees of freedom of the rotor, leading to random movements.[2-3]

To make unidirectional rotation possible, one should include the electronic excited states of the molecule (completely or in a quantum mixing),[4] a possibility which is absent for a classical rotor. This is for example the case of the Feringa et al. molecular motors,[5] which rotate always in the same direction when excited by UV light in solution, bringing the molecules from their electronic ground state ($S_0$) to their excited state ($S_1$).[5]

The intramolecular mechanics of an isolated molecule is quantized. A simple example is a single $NH_3$ molecule in a cold molecular trap, where rotation and vibration spectra can be recorded.[6-7] Instead, using $PF_3$ molecules on a solid surface, leads to gearing effects on the electronic ground state. This was first detected by electron stimulated desorption spectroscopy.[8-9] On a surface, however, the soft quantum mechanical degrees of freedom of both $NH_3$ and $PF_3$ rapidly decohere, and the rotation around their umbrella axis becomes random and semi-classical.[10] A similar mechanical decoherence occurs for the rotation of larger adsorbed molecules like terbutyl-decacyclene on Cu(100).[11] At room temperature, this molecule was the first to show a classical random rotation after manipulation with the tip of a scanning tunneling microscope (STM). This random rotation occurs in the ground electronic state respecting the microscopic reversibility principle.

At low temperature (T = 5 K), a single $O_2$ molecule can rotate step-by-step, but still randomly.[12] W. Ho and co-workers were the first using STM inelastic tunneling electronic excitation to trigger the rotation of a single isolated molecule. Also in that case, the $O_2$ rotational motion is semi-classical, but the energy for the rotation is delivered by the tunneling current producing multiple electron transfer events per second through $O_2$. At each electron transfer event, a quantum mixing of the virtual cationic ground state and the anionic excited electronic states occurs, with a very short femtosecond quantum occupation time.[12] This experiment was followed by several others, where chiral or asymmetric molecules adsorbed on a surface were investigated aiming at reaching unidirectional rotation by inelastic electron tunneling.[2-3, 13-16]

Keeping the STM bias voltage and current as small as possible to avoid any inelastic tunneling effects, molecule-rotors in the ground state were further investigated by increasing the surface temperature, thus observing random rotations.[2] Recently, the unidirectional rotation induced by inelastic tunneling at low temperature was shown to be maintained also at a temperature higher than 5 K.[3] In that case, the structural asymmetry was not provided by the small $C_2H_2$ molecule itself but by the chiral supporting surface.

Understanding the interplay between thermal and electronic excitations for the controlled rotation of a single-molecule rotor is then important for the design of single-molecule machines able to store thermal energy or to produce work. In terms of quantum engineering, this opens fascinating perspectives in the direction of mono-thermal motors.[17]

In this article, we studied the role of thermal energy and electronic excitations for the controlled rotation of a single-molecule rotor on a metal surface. We first separately investigated the starting point of random thermal rotations (by increasing the temperature from T = 5 K) and of unidirectional rotations induced by inelastic tunneling electrons (by STM bias voltage pulses small in voltage and very long in time keeping T = 5 K). Then, we combined tunneling electrons with moderate heating to understand how thermal energy contributes to the one-way rotation.

METHODS

The Au(111) single crystal was cleaned by subsequent cycles of Ar+ sputtering and annealing to 450 °C. 2-(1,3-dimethyl-1H-naphtho[2,3-d]imidazol-3-ium-2-yl) phenolate (DMNI-P) molecules were sublimated from a quartz crucible at T = 225 °C on a clean Au(111) surface held at room temperature under ultra-high vacuum (UHV) conditions (p ≈ 1 × $10^{-10}$ mbar). STM experiments were performed by using a custom-built instrument operating at low temperature under UHV conditions. For thermal excitation measurement, the surface temperature was controlled by a Zener diode (I, V) with a temperature measurement resolution of 0.1 K. The system was stabilized for at least 1 hour in order to achieve thermal equilibrium for each target temperature step. All shown STM images were recorded in constant-current mode with the bias voltage applied to the sample.

Voltage pulses manipulations were performed by positioning the STM tip above the molecule and subsequently ramping up the voltage bias. Both constant current and constant height modes were employed depending on the types of measurements. z(t) or I(t) curves, respectively, were recorded during the pulses, detecting the movement of the molecule. For constant height measurements, the tip–surface distance was calibrated by recording I(z) curves. STM images were recorded before and after the application of the pulse, determining the displacement of the molecule. STM images were taken under non-destructive parameters (e.g., V = 0.2 V, I = 5 pA). The rotation events were collected from the signal of tip height over time, where a sudden jump of signal can be observed. The corresponding yield (events/electron) calculations were done by the average of the population for all fixed voltages and currents (typically n ≥ 10 for each point unless specifically described), and the statistical error was evaluated by standard deviation.

For geometry optimization and reaction path calculations, we used the DFT method as implemented in the CP2K software package (cp2k.org) with the Quickstep module[18]. The Perdew-Burke-Ernzerhof exchange-correlation functional[19], the Goedecker-Teter-Hutter pseudo-potentials[20] and the valence double-ζ basis sets, in combination with the DFT-D2 method of Grimme[21] for van der Waals (vdW) correction were applied. We used 6 layers of gold, where the 3 upper layers were allowed to be relaxed (planar supercell 29.8 x 19.9 Å, vacuum size 40 Å, maximum force 4.5 x 10$^{-5}$ a.u.). The data was analysed, and the images were generated by the PyMOL Molecular Graphics System, Version 2.4 open-source build, Schrödinger, LLC.

The fitting of the action spectrum of Fig. 4c by the theory of Ref.[22] was calculated by:

$$Y = K \left[ \frac{2}{\pi}\left(1 - \frac{\hbar\Omega}{eV}\right)\left(\tan^{-1}\frac{2(eV - \hbar\Omega)}{\sigma} + \tan^{-1}\frac{2\hbar\Omega}{\sigma}\right) + \frac{\sigma}{2\pi eV}\log\frac{((\hbar\Omega)^2 + (\sigma/2)^2)}{((eV - \hbar\Omega)^2 + (\sigma/2)^2)} \right]$$

where $K$ is the overall yield that controls the order of magnitude of the measured yield $Y$, $V$ is the applied bias, $\Omega$ is the angular frequency of the vibrational mode mediating the reaction, and $\sigma$ is the broadening of the reaction threshold. The fitting curves are controlled by basically two parameters; $\Omega$ fixes the threshold of a non-zero yield, and $\sigma$ controls how fast the yield goes from zero to a measurable value.

RESULTS AND DISCUSSION

We selected the DMNI-P molecule-rotor chemisorbed on Au(111) (Figure 1). Compared to other single-molecule rotors,[3, 12, 23] this molecule strongly chemisorbs at low molecular coverage on Au(111), making it impossible to induce any lateral movement on the surface.[15] In the STM tunnel junction, the needed structural asymmetry is carried internally by the molecule-rotor itself. At a temperature of 5 K, the isolated DMNI-P molecules in known to reliably rotate in one direction by a bias voltage of V ≥ 350 mV applied for a few seconds.[15] This specific molecule-rotor therefore represents a well-suitable model system to investigate the onset of the unidirectional rotation by combining thermal and low-voltage tunneling electronic excitations.

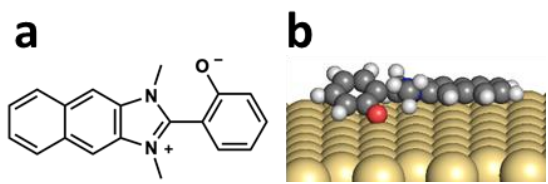

**Figure 1.** (a) Chemical structure and (b) adsorption geometry of the DMNI-P molecule-rotor on the Au(111) surface, calculated by density functional theory.

As presented in Figure 2, at T = 5 K a series of voltage pulses (typically V = 500 mV, I = 250 pA, t = 10 s) produces the controlled unidirectional rotation of the molecule in steps of 60° around the anchoring position at the oxygen atom. Note that the entry port for the electron tunnelling allowing a one-step rotation is at the methyl group on the opposite side of the Au–O bond, where repositioning of the tip after each pulse is necessary. Since DMNI-P is a chiral with the fixed Au-O anchoring point (Figure 2: namely a "left DMNI-P"), we observed opposite rotational direction with right DMNI-P.[15] At these tunneling conditions, a single rotation step of 60° always in the same direction is reproducibly (99% of the cases from 195 pulses on different rotors) observed within 5 s. Such reaction time is extracted plotting the tip height over time during the pulse (z(t), see Figure S2 Supporting Information).

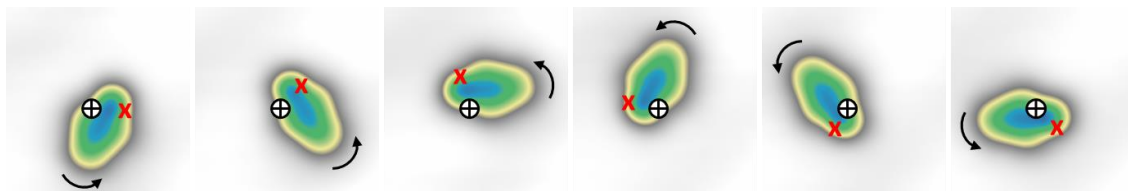

**Figure 2**. STM images sequence of the tip-induced unidirectional rotation of DMNI-P on Au(111). The molecule rotates counterclockwise (CCW) over six stations of 60°. From left to right: STM topography of the same DMNI-P rotor at each step. The position of the STM tip during the voltage pulse is depicted as "x". The black mark indicates the anchoring position. Each voltage pulses were applied under I = 250 pA and V = 0.5 V for 5 s in constant current mode. Images size: 3 nm x 3 nm, I = 5 pA and V = 0.2 V.

**Thermal excitation**

We first consider the thermal heating of the Au(111) surface. We stepwise increased the surface temperature T by the steps of 0.5 – 1 K in order to determine the minimum T triggering a random rotation of the molecule-rotor. For each temperature step, we let the system relax reaching a thermal equilibrium in the STM junction. For STM images we used very low voltage and tunneling current (typically I = 5 pA and V = 10 mV). At these conditions, inelastic tunneling effects involving electronic excited states and tip interactions with the molecule-rotor are negligible. The first rotation event was detected at T = 12 K after waiting 17 hours (See Supplementary Movie 1). When increasing T further, we observed more frequent random rotary motions (Figure 3a and 3b recorded at T = 17 K and T = 23 K respectively, and Supplementary Movie 2 for T = 15 K). At T = 29 K, the STM image recorded during the rotation of the molecule shows a hexagonal shape (Figure 3c) indicating that the molecule-rotor randomly visits the six stable rotational stations during the scanning.

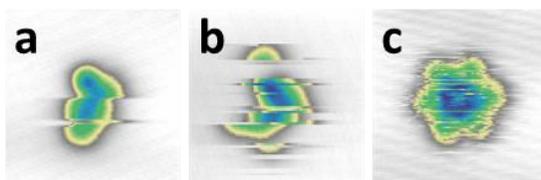

**Figure 3.** Thermal activation of DMNI-P rotation. (a) The molecule starts rotating at T = 17 K, (b) while it becomes faster at 23 K. (c) At 29 K it appears as hexagonal shape since it rotates faster than the scanning of STM image. STM images: size 4 nm x 4 nm, (a) I = 5 pA, V = 100mV, (b) I = 5 pA, V = 200 mV, and (c) I = 5 pA, V = 10 mV, scanning speed: 17.3 s/image. The thermal excited rotation frequency shows an Arrhenius behavior, resulting in an energy barrier of $E_{ex}$ = 29.2 ± 4.7 meV.

The observed thermal-induced random rotation shows an Arrhenius behavior indicating a rotation energy barrier of $E_{ex}$ = (29 ± 5) meV (Figure S3, Supporting Information), significantly higher than the value obtained for the shorter DMBI-P molecule-rotor with a phenyl less,[2] where a rotation energy barrier $E_{ex}$ = (5 ± 2) meV was observed. The extra phenyl increases the van der Waals interactions with the Au(111) surface, as confirmed by DFT calculations with vdW corrections (see Supporting Information).

In agreement with the microscopic reversibility principle, the experiment confirms that a purely thermal heating of the DMNI-P molecule-rotor does not induce a unidirectional rotation, but only a random one. It is also interesting to note that the anchoring of the molecule-rotor on the surface survives at least up to 77 K, where random and very fast oscillations are observed (Supplementary

Figure S4). Even at this relatively high surface temperature, the rotational dynamics remains in the electronic ground state, where only 6 stable stations can be observed on Au(111).

**Electronic excitation at low temperature**

In the second part of the experiment, we kept the temperature constant at T = 5 K, aiming at determining the smallest possible electronic excitation inducing a directional one-step rotation of 60°. By progressively reducing the STM bias voltage, we measured how the time needed to observe a one-step unidirectional rotation event extends from a few seconds to many hours (Figure 4a and 4b). The experiment was systematically repeated on the same molecule (shown in the inset of Figure 4a) after decreasing the bias voltage and increasing the pulsing time duration (always keeping the same tunneling current I = 700 pA). At 150 mV, the first one-step directed rotation event was observed while keeping the STM tip apex at the same position on the molecule-rotor for 35 950 s (around 10 h) in a 12 h pulse (applied at the position of the red cross in the inset of Figure 4a). The lateral thermal drift is negligible in our experimental setup.

After 139 pulses, we plotted the response time as a function of the bias voltage (Figure 4a). Each response time is measured on the corresponding z(t) curve (like in Figure 4b). The obtained experimental yield, or action spectrum[22] (Y = R(I/e)), is plotted in Figure 4c. For V > 400 mV, we confirm the nearly constant yield of about 5 x $10^{-10}$ rotations/electron previously reported,[15] which is assigned to the onset of the C-H stretch vibrational mode at 370 meV. This vibration is the principal energy entry port to trigger a one-step rotation event in the higher bias voltage range. At lower bias voltage, we observe a sharp drop of the yield between 330 and 400 mV, while another nearly constant region appears at 200 – 300 mV. For V < 200 mV, a rapid decrease of yield is again observed, reaching the low yield of 6 x $10^{-15}$ rotations/electron at 150 mV.

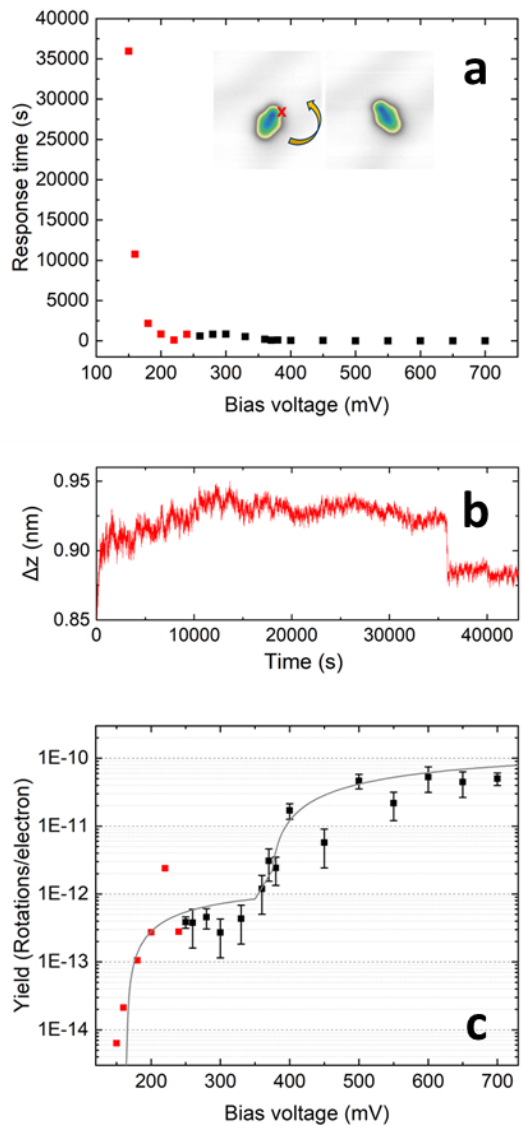

**Figure 4.** (a) Response time for a unidirectional rotation over the applied bias voltage. Inset shows a one-step directional rotation of DMNI-P, where the red cross depicts the position of the tip during voltage pulse. (b) Relative tip height versus time Δz(t) measured during the pulse. A rotation event was observed after 35 950 s in a 12 h pulse (I = 700pA, V = 150 mV). (c) Yield versus bias voltage for the rotor at pulses I = 700 pA. Due to the long waiting time, no statistical analysis was possible and only one (red-marked) data point was recorded in (b). The data have been fitted by the action spectra function of Ref. 19 for two modes, one at 165 mV and one at 370 mV.

By fitting the data (continuous line in Figure 4c), there are two thresholds, one at 165 meV and the other at 370 meV, corresponding to a large density of DMNI-P deformation mode (C-C stretch manifold) and the above-mentioned C-H stretch mode of DMNI-P, respectively. Those modes contribute to the unidirectional rotation, providing the rotational energy inelastically.

**Thermal with inelastic electronic excitation**

To investigate the interplay between thermal energy and tunneling electronic excitations, we determined the range of temperatures and STM bias voltage pulses where both thermal heating and electronic excitation contribute to the rotation. We therefore repeated the above rotation experiments by applying STM voltage pulses (50 mV < V < 500 mV) increasing progressively the temperature in the interval 5 K < T ≤ 17 K. In such conditions, we observed a faster rate of unidirectional rotation for T > 5 K while applying voltage pulses.

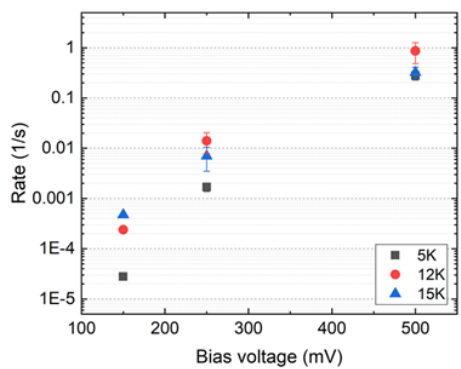

**Figure 5.** Interplay between thermal and inelastic electronic excitations for unidirectional rotation. Rotation rate dependence on the bias voltage at various temperatures (5, 12, and 15 K) at I = 700 pA.

In Figure 5, the rate of the unidirectional rotation (1/reaction time) depending on the voltage is plotted for three different temperatures (all data points measured at I = 700 pA). Increasing the temperature leads to a faster response time for unidirectional rotation. This effect is diminished by increasing the voltage, becoming negligible for V > 500 mV. At this energy, the C-H vibrational mode is excited and the inelastic electronic excitation becomes the dominating energy source for a one-way rotation, thus overcoming the STM junction temperature contribution. Action spectra taken at T = 12 K and 15 K are shown in Supplementary Figure S5.

In Figure 6, we summarize the mechanical response of a single DMNI-P molecule-rotor by the combination of thermal and tunneling electronic contributions. On the x,y base plane, the applied voltage and temperature are indicated. The blue dots correspond to observed unidirectional rotation, and the orange dots to the (T, V) values where we observed random rotations. On the x-z vertical curve, the yield curve is sketched from Figure 4c for the inelastic electronic excitation at T = 5 K, illustrating the two main thresholds. According to Figure 4c, three regions are determined by the excitation energy threshold of soft mechanical and the C-H stretch modes at 165 meV and 370 meV, respectively.

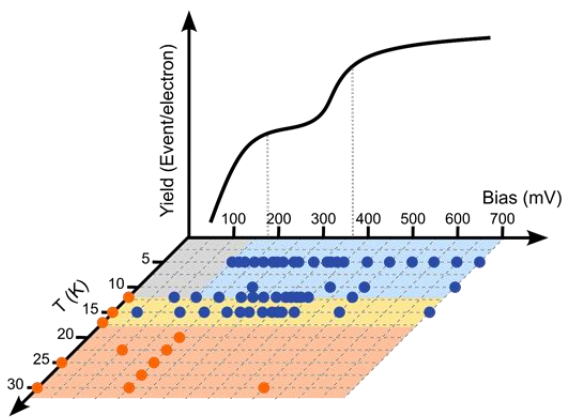

**Figure 6**. Summary of combining thermal and electronic excitations. Plot of yield over bias voltage and temperatures for rotation. The orange dots indicate the conditions for random rotation, the blue ones the unidirectional rotations. The yield curve is sketched from Figure 4c.

In Figure 6, we can identify four different (T, V) rotation regimes, indicated with different colors in Figure 6, and discussed in the following:

(1) The small grey area indicates the very low voltages and surface temperature regimes, where no rotation events have been observed, or experiments were not performed. In those conditions we do not expect the creation of a rotation minimum energy path (random or not) on the ground state potential energy surface.

(2) The light blue area corresponds to the (T, V) values between 5 K < T < 12 K where a step-by-step unidirectional rotation takes place by applying a bias voltage (150 mV < V < 700 mV). In this area, the inelastic tunneling electrons provide energy to the molecule-rotor via vibrational modes. At each electron transfer event, such tunneling electrons virtually occupy both cationic ground $S_0+(\theta)$ and anionic excited $S_1-(\theta)$ electronic states, where $\theta$ is the angle of rotation.[2] The occupation is virtual because the $S_0+$ and $S_1-$ tunneling resonances are more than 2 eV far from the bias voltage of the C-H stretching excitation (see Supplementary Figure S6). In these conditions, the vibrational energy is able to trigger a dynamical path on the potential energy surface $S_1-(\theta)$ to pass over the rotation barrier, without contradicting the microscopic reversibility principle. The needed asymmetry of the $S_1-(\theta)$ surface respect to the $S_0+(\theta)$ is coming from the chirality of the molecule and by its charge separation in the electric field of the STM junction. At small bias voltage, the long response time can be explained by the competition between the intramolecular vibration energy redistribution and the energy relaxation towards the Au(111) surface.

(3) In the yellow area, both blue and orange points are present, *i.e.,* both unidirectional and random-direction rotations are possible. When reducing the bias voltage to nearly 0 V, the molecule rotates randomly. Interestingly, increasing the surface temperature (12 K < T < 20 K) with a bias voltage 50 mV ≤ V ≤ 250 mV keeps the rotation unidirectional, thus significantly reducing the rotation response time (Figure 5). In this narrow (T, V) range, the inelastic tunneling effect is weak with a low probability of exciting low-energy vibrational modes at 165 meV and below (Figure 4c). As already mentioned, the tunneling current can be described by a quantum mixing of the $S_0+(\theta)$ and $S_1-(\theta)$ virtual electronic configurations. Therefore, at each electron transfer event, part of the thermal energy coming from the surface is redistributed intramolecularly and trigger a dynamical path on the $S_1-(\theta)$ potential energy surface. This means that the rotational barrier on $S_1-(\theta)$ can be overcome, and a one-step of unidirectional rotation happens similarly as discussed above for inelastic effects. In other words, each electron transfer event "projects" part of the thermal energy captured by the molecule-rotor on the $S_1-(\theta)$ state, practically sharing it between $S_0+(\theta)$ and $S_1-(\theta)$. This is a quantum motor as presently discussed in quantum information theory.[17] The thermal contribution to the unidirectional rotation diminishes at higher voltages (Figure 5), indicating that the opening of the inelastic C-H stretch channel is taking the lead to project energy to the $S_1-(\theta)$ state, making the thermal contribution negligible.

(4) In the orange area (T > 20 K and independently of the bias voltage 0 < V < 700 mV) the rotation is non-directional. The surface temperature is now too large to take advantage from the topology of the $S_1-(\theta)$ potential energy surface relative to $S_0+(\theta)$ described above, and the rotation becomes a stochastic process.

CONCLUSIONS

In conclusion, we have investigated the unidirectional rotation of chemisorbed DMNI-P molecule-rotors on the Au(111) surface induced by thermal and tunneling electronic excitations.

While in a well identified voltage and thermal range the induced rotation can be clearly separated as due to thermal effect (random rotation) or electronic inelastic excitations (unidirectional rotation), there is a narrow area of voltages and temperatures where the surface thermal excitation enhances the unidirectional rotational rate. This effect can be rationalized describing the electron transfer events during tunneling by a quantum mixing of virtual electronic states. During each of those events, part of the thermal energy captured from the surface is distributed among the electronic states of the molecule-rotor, increasing the probability of overcoming the rotational energy barrier. In this specific voltage range, the probability of inelastic excitation of vibrational modes is small, and a unidirectional rotation of the molecule-rotor is observed even without involving inelastic tunneling effect, opening new perspectives for the development of quantum motors.

ASSOCIATED CONTENT

**Supporting Information.** The following files are available free of charge:

- A supporting information PDF file is containing further experimental results and supplementary calculations.

- A mp4 movie (movie1.mp4) showing the random movement at T = 12 K. The STM images are repeated every 15 minutes. All STM images were taken under I = 5 pA and V = 100 mV, image size 30 nm x 30 nm.

- A mp4 movie (movie2.mp4) showing the random movement at T = 15 K. The STM images are repeated every 15 minutes. All STM images were taken under I = 5 pA and V = 50 mV, image size 30 nm x 30 nm.

AUTHOR INFORMATION

**Corresponding Authors.** *Email: francesca.moresco@tu-dresden.de

**Author Contributions.** The manuscript was written through contributions of all authors. All authors have given approval to the final version of the manuscript.

ACKNOWLEDGMENT

This work was funded by the European Union. Views and opinions expressed are however those of the authors only and do not necessarily reflect those of the European Union. Neither the European Union nor the granting authority can be held responsible for them.

This work has received funding from then European Innovation Council (EIC) under the project ESiM (grant agreement No 101046364) and the Horizon 2020 research and innovation program under the project MEMO, grant agreement no. 766864.

Support by the German Research Foundation (DFG) by the DFG Project 43234550, the Collaborative Research Centre (CRC) 1415, and the Initiative and Networking Fund of the German Helmholtz Association, Helmholtz International Research School for Nanoelectronic Networks NanoNet (VH-KO-606) is gratefully acknowledged.

F.L. thanks the Fonds der Chemischen Industrie (FCI) for a Liebig Fellowship and C.J. the WPI MANA project for financial support.

R.R. and N.L. acknowledge financial support from the Spanish State Research Agency Grant PID2021-127917NB-I00 funded by MCIN/AEI/10.13039/501100011033 and by "ERDF A way of making Europe".

REFERENCES

1. Tolman, R. C. The Principle of Microscopic Reversibility. *Proc Natl Acad Sci U S A* **1925,** *11*, 436-439.
2. Eisenhut, F.; Kühne, T.; Monsalve, J.; Srivastava, S.; Ryndyk, D. A.; Cuniberti, G.; Aiboudi, O.; Lissel, F.; Zobač, V.; Robles, R. *et al.* One-Way Rotation of a Chemically Anchored Single Molecule-Rotor. *Nanoscale* **2021,** *13*, 16077-16083.
3. Stolz, S.; Gröning, O.; Prinz, J.; Brune, H.; Widmer, R. Molecular Motor Crossing the Frontier of Classical to Quantum Tunneling Motion. *Proc. Natl. Acad. Sci.* **2020,** *117*, 14838-14842.
4. Joachim, C. The Driving Power of the Quantum Superposition Principle for Molecule-Machines. *J. Phys.: Condens. Matter* **2006,** *18*, S1935.
5. Koumura, N.; Zijlstra, R. W. J.; van Delden, R. A.; Harada, N.; Feringa, B. L. Light-Driven Monodirectional Molecular Rotor. *Nature* **1999,** *401*, 152-155.
6. Bethlem, H. L.; Berden, G.; Crompvoets, F. M. H.; Jongma, R. T.; van Roij, A. J. A.; Meijer, G. Electrostatic Trapping of Ammonia Molecules. *Nature* **2000,** *406*, 491-494.
7. van Veldhoven, J.; Küpper, J.; Bethlem, H. L.; Sartakov, B.; van Roij, A. J. A.; Meijer, G. Decelerated Molecular Beams for High-Resolution Spectroscopy. *Eur. Phys. J. D* **2004,** *31*, 337-349.
8. Zwanzig, R. Rotational Relaxation in a Gear Network. *The Journal of Chemical Physics* **1987,** *87*, 4870-4872.
9. Alvey, M. D.; Jr., J. T. Y.; Uram, K. J. The Direct Observation of Hindered Rotation of a Chemisorbed Molecule: Pf3 on Ni(111). *The Journal of Chemical Physics* **1987,** *87*, 7221-7228.
10. Klauber, C.; Alvey, M. D.; Yates, J. T. Evidence for Chemisorption Site Selection Based on an Electron-Donor Mechanism. *Chem. Phys. Lett.* **1984,** *106*, 477-481.
11. Gimzewski, J. K.; Joachim, C.; Schlittler, R. R.; Langlais, V.; Tang, H.; Johannsen, I. Rotation of a Single Molecule within a Supramolecular Bearing. *Science* **1998,** *281*, 531-533.
12. Stipe, B. C.; Rezaei, M. A.; Ho, W. Inducing and Viewing the Rotational Motion of a Single Molecule. *Science* **1998,** *279*, 1907-1909.
13. Tierney, H. L.; Murphy, C. J.; Jewell, A. D.; Baber, A. E.; Iski, E. V.; Khodaverdian, H. Y.; McGuire, A. F.; Klebanov, N.; Sykes, E. C. H. Experimental Demonstration of a Single-Molecule Electric Motor. *Nat. Nanotechnol.* **2011,** *6*, 625-629.
14. Perera, U. G. E.; Ample, F.; Kersell, H.; Zhang, Y.; Vives, G.; Echeverria, J.; Grisolia, M.; Rapenne, G.; Joachim, C.; Hla, S. W. Controlled Clockwise and Anticlockwise Rotational Switching of a Molecular Motor. *Nat. Nanotechnol.* **2013,** *8*, 46-51.
15. Au-Yeung, K. H.; Sarkar, S.; Kühne, T.; Aiboudi, O.; Ryndyk, D. A.; Robles, R.; Lorente, N.; Lissel, F.; Joachim, C.; Moresco, F. A Nanocar and Rotor in One Molecule. *ACS Nano* **2023,** *17*, 3128-3134.
16. Jasper-Toennies, T.; Gruber, M.; Johannsen, S.; Frederiksen, T.; Garcia-Lekue, A.; Jäkel, T.; Roehricht, F.; Herges, R.; Berndt, R. Rotation of Ethoxy and Ethyl Moieties on a Molecular Platform on Au(111). *ACS Nano* **2020,** *14*, 3907-3916.
17. Scully, M. O.; Zubairy, M. S.; Agarwal, G. S.; Walther, H. Extracting Work from a Single Heat Bath Via Vanishing Quantum Coherence. *Science* **2003,** *299*, 862-864.
18. VandeVondele, J.; Krack, M.; Mohamed, F.; Parrinello, M.; Chassaing, T.; Hutter, J. Quickstep: Fast and Accurate Density Functional Calculations Using a Mixed Gaussian and Plane Waves Approach. *Comput. Phys. Commun.* **2005,** *167*, 103-128.

19.     Perdew, J. P.; Burke, K.; Ernzerhof, M. Generalized Gradient Approximation Made Simple. *Phys. Rev. Lett.* **1996,** *77*, 3865-3868.
20.     Goedecker, S.; Teter, M.; Hutter, J. Separable Dual-Space Gaussian Pseudopotentials. *Phys. Rev. B* **1996,** *54*, 1703-1710.
21.     Grimme, S.; Antony, J.; Ehrlich, S.; Krieg, H. A Consistent and Accurate Ab Initio Parametrization of Density Functional Dispersion Correction (Dft-D) for the 94 Elements H-Pu. *J. Chem. Phys.* **2010,** *132*, 154104.
22.     Kim, Y.; Motobayashi, K.; Frederiksen, T.; Ueba, H.; Kawai, M. Action Spectroscopy for Single-Molecule Reactions – Experiments and Theory. *Prog. Surf. Sci.* **2015,** *90*, 85-143.
23.     Simpson, G. J.; García-López, V.; Daniel Boese, A.; Tour, J. M.; Grill, L. How to Control Single-Molecule Rotation. *Nat. Commun.* **2019,** *10*, 4631.

TABLE OF CONTENTS GRAPHIC

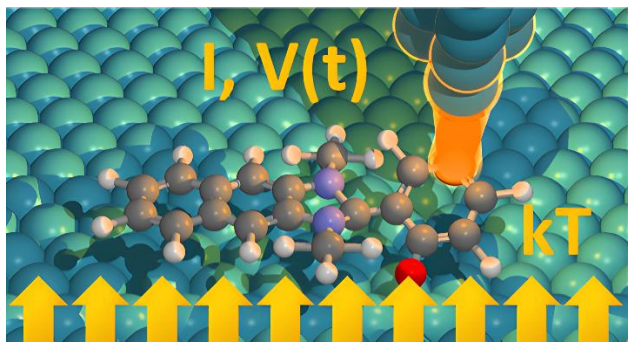